\documentclass[useAMS,usenatbib]{mn2e}
\usepackage{graphicx}
\usepackage{amsmath}

\usepackage{soul}

\DeclareGraphicsExtensions{.pdf, .jpg, .png}

\graphicspath{{./figures/}}

\title[Stirring N-body systems (Part 2)]{Stirring N-body systems II: Necessary conditions for the dark matter attractor}
\author[Jeremy A. Barber, Hongsheng Zhao and Steen Hansen]{Jeremy A. Barber$^{1}$\thanks{E-mail:
jab22@st-andrews.ac.uk (JAB); hz4@st-andrews.ac.uk (HZ); hansen@dark-cosmology.dk (SH)}, Hongsheng Zhao$^{1}$ and Steen H. Hansen$^{2}$\\
$^{1}$SUPA, University of St Andrews, North Haugh, St Andrews, Fife, KY16 9SS, UK\\
$^{2}$Dark Cosmology Centre, Niels Bohr Institute, University of Copenhagen, Juliane Maries Vej 30, 2100 Copenhagen, Denmark
}
\begin{document}

\date{Accepted ----. Received ----}

\pagerange{\pageref{firstpage}--\pageref{lastpage}} \pubyear{2014}
\maketitle

\label{firstpage}

\begin{abstract}
We study the evolution of the phase-space of collisionless
N-body systems under repeated stirrings or perturbations, which has
been shown to lead to a convergence towards a limited group of end states.
This so-called attractor was previously shown to be independent of the
initial system and environmental conditions. However the fundamental
reason for its appearance is still unclear. It has been suggested that
the origin of the attractor may be either radial infall (RI), 
the radial orbit instability (ROI), 
or energy exchange which, for instance, happens during violent relaxation. 
Here we examine the effects of a set of controlled perturbations, referred to as `kicks',
  which act in addition to the standard collisionless dynamics by
  allowing pre-specified instantaneous perturbations in
  phase-space. We first demonstrate that the attractor persists in the absence of RI and ROI by forcing the system to expand. We then consider radial velocity kicks in a rigid
potential and isotropic velocity kicks, since there are no energy
exchanges in these two recipes of kicks. We find that these kicks do
not lead to the attractor, indicating that the energy exchange in a dynamic potential is
indeed the physical mechanism responsible for the attractor.
\end{abstract}


\section{Introduction}

It is known that N-body simulations tend to yield a narrow
range of density profiles for stellar bulges or dark halos \citep[and
references therein]{Dubinski1991}. These are often parameterised using
sub-families of profiles of \citet{Zhao1996}such as the
Hernquist profile \citep{Hernquist1990} for stellar spheroids or the
NFW profile \citep{Navarro1996} for halos. These density profiles and
corresponding velocity anisotropy profiles are a result of the energy
and angular momentum distribution of particles in the simulation
\citep{Spergel1992}, especially how particles exchange their energy
and angular momentum, as well as the dynamical friction/tidal
disruption of subclumps and the absence/existence of an expanding
cosmic background \citep{Syer1998, Taylor2001}.

Any N-body code must have its physical laws for such things programmed into
it \emph{a priori}, an act which implies that those laws and their
implications are comparatively well understood. This paper is part of
an ongoing investigation into an \emph{unexpected} result, namely
the identification of an attractor in the phase space of N-body
systems, as discussed in \citet{Hansen2010}. The attractor is a
single-valued relationship between parameters of the Jeans equation; 
parameters which could have been entirely independent.

The formulation of the Jeans equation that we make use of is:

\begin{equation}
\label{eqn:jeans}
v_c^2=-\sigma_r^2(\gamma+\kappa+2\beta)
\end{equation}

where $v_c$ is the circular speed, $\sigma_r^2$ is the radial velocity
dispersion, $\gamma$ and $\kappa$ are the gradients of $log(\rho)$ and
$log(\sigma_r^2)$ respectively with respect to $log(r)$, and finally
$\beta$ is the velocity anisotropy.

The `attractor' relationship appears if an arbitrary system was repeatedly randomly perturbed
away from equilibrium and then allowed to settle again. The perturbations consist of an alternating 
cycle of perturbations, referred to as `kicks', with a subsequent period of relaxation to a new
equilibrium state, referred to as the `flow'.

The kicks are a set of controlled, artificial perturbations 
which act in addition to the standard collisionless dynamics by
allowing pre-specified perturbations in
phase-space. These perturbations were often constructed to be instantaneous and
in most previous studies these kicks allowed energy exchange amongst a set of particles, even though
these particles have negligible two-body energy exchange. 

A system perturbed with such repeated kick-flows demonstrates a strong tendency to equilibrate along solutions, regardless of the
initial characteristics of the perturbed system \citep{Barber2012, HansenOct}, as follows:

\begin{equation}
\label{eqn:attractor}
\beta=\frac{-0.15\gamma-0.85\kappa}{(1+(-0.15\gamma-0.85\kappa)^3)^{\frac{1}{3}}} \, .
\end{equation}

This is not the first time that relations and constraints have been discovered in components of the Jeans equation, nor is it the first time specific relations between $\beta$ and $\gamma$ have been identified \citep{Hansen2004,An2006, Ciotti2010}.

Our previous paper primarily focused on the bulk properties of the
initial system and the statistics of the perturbation schemes being
used. In this paper we focus on the fundamental requirements of the
emergence of this attractor phenomenon.

This is particularly relevant for observations as, depending on the
physics responsible for the attractor, it may be either be irrelevant for cosmological structure or it may be of
fundamental importance to all equilibrated structures. As we
will explain, the attractor may appear due to physics which 
happens during each merger throughout the history of structure
formation. We will demonstrate that
this is the case, rendering the attractor potentially very 
important for the equilibrated part of all cosmological structures, 
in agreement with the results of large cosmological simulations \citep{Ludlow2010}.

This paper focuses on the origin of the attractor in spherically
averaged quantities in an effort to find out what the driving factors
for the convergence are. For comparison two recent papers, \citet{HansenMay, HansenOct}, discuss
how the attractor does appear in the spherically averaged
characteristics of many halos but becomes more complicated when the
merger history of the object is considered. Those works, using
radially-aligned, conical bins rather than spherical averages, find
deviations from the attractor along preferred axes defined by the vectors along which past mergers had taken place. However, spherically
averaged properties still follow the attractor in most cases.

As this work uses the same analysis pipeline as our previous paper, \citet{Barber2012} (hereafter B12), a detailed description of the method can be found there. In summary we use NMODY, a particle-mesh code developed for use with Modified Newtonian Dynamics \citep{Ciotti2006}, although it is used only in its purely Newtonian mode here.
The systems used throughout this paper are Plummer spheres of scale radius 0.05$\,$kpc, containing $5\times 10^8$ $M_{\sun}$ in $750,000$ particles. Plummer spheres were chosen as they are formally unrelated to the NFW profile and are easy to create with varying anisotropies using the methods of \citet{Gerhard1991}, based on using a particular distribution function into two independent functions that represent the distribution of energy and angular momentum respectively.

The paper is structured as
follows: \S2 will be on the effect of collision and resolution, \S3 will be on radial infall and the radial orbit instability, 
\S4 gives two recipes to avoid the attractor and \S5 concludes.

\section{Impact of numerical resolution effects}
We begin by examining the possibility of a connection between our results and the numerical resolution of our simulations; the suggestion being that our conclusions were heavily influenced by numerical artifacts rather than by the physics of the system. Of specific concern was whether the behaviour we were describing could be caused by collisional relaxation and would thus be governed by the softening length of the simulation.

NMODY uses a self-consistent field (SCF) method similar to that described in \citet{HernquistOstriker1992} \emph{i.e.} it describes the potential and density by expanding them into a series of terms in some basis functions. \cite{Ciotti2006} is mainly concerned with demonstrating accurate recovery of MONDian potential-density pairs via this method and thus does not spend time looking at the impact of resolution explicitly. However, the two methods are sufficiently similar that informative results can still be found by examining the impact of numerical resolution in the \cite{HernquistOstriker1992} method.

This work showed that for a variety of initial density models the relative importance of each subsequent term in the basis series decreases exponentially for Plummer models such as ours, providing better than $1\%$ accuracy, in terms of orbit conservation, when using around $5$ terms in the series. It is noted that cored models, such as a Plummer sphere, can be particularly well described by this kind of expansion method if the basis functions are chosen appropriately.

Having demonstrated the accuracy of the method, \cite{HernquistOstriker1992} examines the emergence of collisional relaxation in the such a simulation. As there is no explicit softening length used in the method they note that an SCF code should not suffer from limitations on spatial resolution and can, in principle, resolve much steeper density gradients than other methods. Overall, from the conclusion of this paper, we would expect that a code such as NMODY would be efficient at suppressing relaxation noise as only a handful of basis terms are required to provide ample spatial resolution for the simulation.

\begin{figure}
\includegraphics[width=84mm]{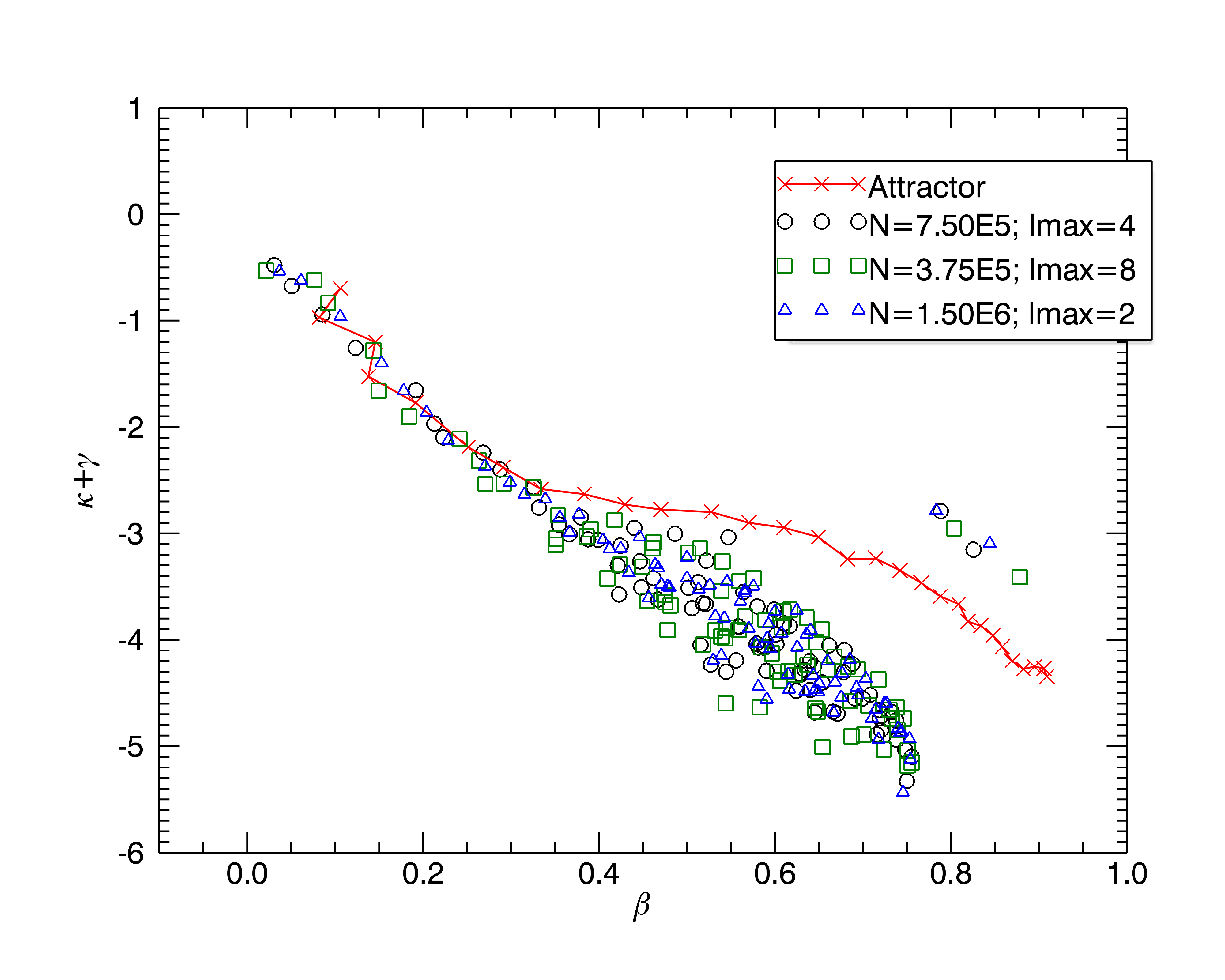}
\caption{\label{fig:convergencetest} Attractor plot demonstrating that there are negligible two-body interactions. It shows the state of three simulations approximately one third (in terms of elapsed time) of the way towards convergence with the attractor (red line). All three use the violent exchange algorithm outlined in B12 with scaling factor $0.5$ and $3T_{dyn}$ flow time per kick. Our benchmark simulation from B12 (black circles) overlaps comfortably with both the rougher (green squares) and smoother (blue triangles) bins of the newer simulations demonstrating a very similar rate of convergence. This shows that the evolution of our systems towards the attractor, which is where they eventually rest, is unaffected by the resolution of the simulations and, consequently, by collisional effects.}
\end{figure}

With this in mind, to explicitly examine the impact of smoothing lengths and resolution on our simulations, several supplementary simulations were run that fell into one of two categories; rougher or smoother models, \emph{cf.} figure \ref{fig:convergencetest}.

Rougher models lowered the resolution of the model in two ways. Firstly, when the code developed the spherical harmonics that described the potential it used twice the number of terms in the series, making the potential more variable on shorter scales and thus magnifying the effect of short distance interactions. This approach was used rather than simply changing the smoothing length because, as previously discussed, NMODY does not support the direct selection of a desired smoothing length due to the self-consistent field scheme.

Secondly, the initial conditions modelled the same systems as in B12, but now using half the number of particles. This preserved the dynamical timescale of the system while also making the particle distribution noisier. Overall, we would expect the combination of these two effects to emphasise any effects from collisions and, if they are driving the attractor, to lead to the attractor faster.

Smoother simulations use the same reasoning except they halved the number of terms in the harmonics, effectively smoothing out perturbations on short length scales, and had double the number of particles modelling the system which should smooth the distribution overall. If artificial numerical collisions were governing the attractor, then the de-emphasis of short scale interactions and the smoother particle distribution should suppress the effect.

Overall, these rough/smooth schemes allow us to control the resolution and susceptibility of the system to collisional effects and short-scale interactions while retaining comparable simulations that, as a practical benefit, require comparable amounts of processing time to yield results.

To demonstrate this, we present figure \ref{fig:convergencetest} which shows the state of three comparable simulations 10 kick-flow cycles (10 perturbing events spaced evenly throughout a total simulation time of 30 dynamical times). This is approximately one third of the time required for the systems to reach the attractor, given the chosen magnitude of the perturbation according to B12, and demonstrates that the systems are indistinguishable from each other in terms of the parameter space they occupy. Our simulations eventually end up lying in the parameter space of the attractor in a manner indistinguishable from the results from B12. In particular we establish that the behaviour is preserved down to the speed at which the convergence occurs.

The same rougher/smoother dichotomic scheme was applied to a new perturbation method that performs systematic alterations to the system's velocity anisotropy profile. This new perturbation method will be explained in detail in \S4.1 and it is mentioned here only insofar as to make clear that it also appears unaffected by alterations to the smoothing of the potential.

In summary, the attractor effect is demonstrated to progress the same regardless of the number density of the system or how accurately the simulation models short-scale behaviour and is present in simulations that use different codes (\citet{Hansen2010} used GADGET-2 which has a different architecture to NMODY) to solve for the particles motions. Accordingly, it is not thought that the attractor shares any significant causal link to collisional relaxation or any effect deriving from two-body interactions.

\section{Ruling out Radial Infall by adding energy}

A characteristic feature of the previous simulations from B12 was a significant amount of radial infall (RI) whereby systems would collapse into more radially anisotropic systems as they were kicked. The fact that all the simulations shared this common mechanism raises the question of whether or not RI is a contributing factor to the attractor. 

\subsection{Algorithm for avoiding infall}

To examine whether or not RI is necessary for the attractor we now use a variation of the scheme from
B12 whereby random numbers were used to scale the three cartesian
velocity vectors of the system's particles, however it dispenses with
the now redundant routines for assessing energy conservation for
reasons that will be explained shortly. The scheme is now as follows:

\begin{itemize}
\item Set up a series of radial bins. We chose to create bins defined to contain 5,000 particles.
\item For each particle in each bin we examine each of the three
  orthogonal velocity vectors and multiply each by a random number $f$
  drawn from a uniform distribution centered around 1.5
  e.g. $1.0<f<2.0$. This instantaneous perturbation is referred to as the `kick'
and $f$ can be called the `kick scale factor'. That
  the distribution is \emph{not} centred on unity is what lends
  \emph{this} scheme its desired asymmetry, as any given velocity
  component will be at least as big as it was prior to this scaling.
\item Derive a dynamical timescale for the system
\begin{equation}
\label{eqn:time}
t_{dyn}=\sqrt{\frac{1}{G\rho}}\text{ where }\rho=\frac{0.95\times M_{tot}}{\frac{4}{3}\pi{r_{95\%}^3}}
\end{equation}
where we are using the $95^{th}$ mass percentile as a representative distance for the system. For our initial systems this is equivalent to approximately 3 scale radii and is identical to the previous method.
\item The system is then left to evolve in an N-body simulator for 3 dynamical timescales. This `flow' period allows the system to relax and find a new equilibrium. If we were to apply another kick too soon then the impact of the second kick would be indistinguishable from that of the first.
\item Repeat the cycle as needed.
\end{itemize}

This kick can only be applied a finite
number of times before a significant number of particles become
unbound from the system. After a
large amount of particles become unbound they will no longer interact
with each other and the system ceases to have a meaningful dynamical
time. We find that for this particular kick this effect starts to dominate around the fifth kick
cycle by which point the outermost $50$ bins, \emph{i.e.} $250,000$ particles
or a third of the entire system, have become entirely unbound from the
structure. At this point the simulation is manually halted as the divergent behaviour of the dynamical timescale becomes insurmountable as well as increasingly physically meaningless.

\subsection{Relation between RI and Radial Orbit Instability}

The link between the perturbation used in B12 and RI was in how the system conserved energy. The perturbation in B12 performed the scaling on the components of velocity $v_{x,y,z}$ but then proceeded to conserve energy in the form of $v^2$. This meant that the conservation was asymmetrical compared to the perturbation as $|(v+\delta v)^2-v^2| > |(v-\delta v)^2-v^2|$ unlike the \emph{perturbation} where $|(v+\delta v)-v| = |(v-\delta v)-v|$ \emph{i.e.} the particles that had their overall velocity increased were contributing more to the kinetic energy of the bin than was being removed by particles which had their velocities decreased by the same amount, leading to an overall increase in energy. The energy conservation code worked on all particles equally, so most particles in a bin ended up losing energy overall to compensate for the small fraction of particles which got large velocity increases and thus significant energy increases. This sudden removal of energy resulted in the compaction of the system and placed many more particles on more radial, infalling orbits.

The main concern was that if RI proved to be necessary for the attractor, and the algorithm was artificially inducing such an infall, then the attractor may just be an artifact of the perturbation scheme. This was compounded by the fact that the radial orbit instability (ROI) displays several convergent behaviours in the parameter spaces of $\beta$ and $\gamma$, as outlined in many papers over the years \citep{Huss1999,Barnes2005,HansenMoore2006,Hansen2006,Macmillan2006,Bellovary2008,Lapi2011}, which bear some noticeable similarities to the attractor.

ROI refers to the unstable nature of orbits in initially spherically symmetric systems which have a large population of their
particles on highly radial orbits. Systems set up in this way will 
depart from spherical equilibrium and will eventually become triaxial systems  
\citep{Antonov1973}. This behaviour was seen in the majority of the
simulations from B12, only noticeably milder than one would expect
from a system governed by ROI due to the mild nature of the
perturbation. It was therefore suggested that the attractor was being driven by the statistical effects of the kick that caused RI. The increased amount of radial orbits could then lead to ROI which would slowly dominate the system giving rise to the convergent behaviours in our parameter space that we called the attractor.

\subsection{Result}

We therefore wished to define an algorithm
that could rule-out, or confirm, RI as a
contributing mechanism. To that end we designed the simple kick outlined above, based on the same algorithm of random numbers as in B12. The key difference is that the kick is now asymmetrical, only
ever adding energy to a bin, never removing it, and we do not enforce
any kind of energy conservation after the kick. The idea is that the
system will expand as a result of the added energy and is thus not placing more particles on radial, inwards orbits, preventing
a collapsing state, and thus not triggering either RI or ROI.

If the system did not evolve towards the attractor or evolved in
a completely different manner now that any kind of infall was being prevented then
that would suggest that RI was an important, necessary factor.

\begin{figure}
\includegraphics[width=84mm]{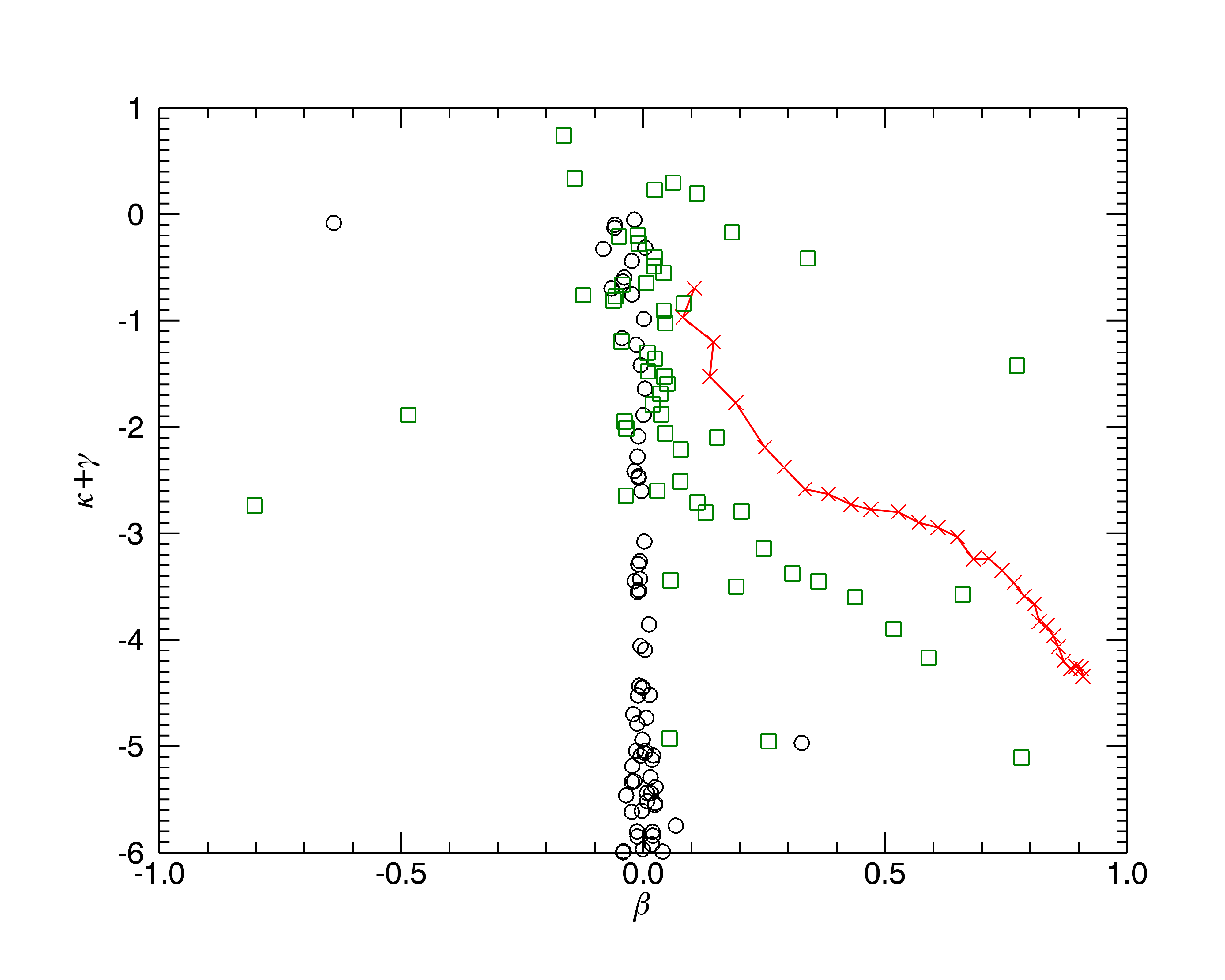}
\caption{\label{fig:biaskick} Plot showing the system's progress
 towards the attractor using the energy-adding kick. Each point represents a mass bin of $5,000$ particles and the red line is the
 position of the attractor. Black
 circles show bins from the initial conditions whilst green squares
 are likewise the state \emph{after 5 kicks}.}
\end{figure}

Figure \ref{fig:biaskick} shows a plot of the attractor for this
kick. The open shapes are bins of $5,000$ particles each and the
red line is data from \citet{Hansen2010} that marks the position of
the attractor in the parameter space. The system shows, despite the
unrealistic kick, significant evolution towards the same space
as the attractor; from the black circles to the green squares. It is not
sitting directly on the attractor but this is, as discussed
previously, because of the amount of unbound material causing the simulation to end prematurely.
The system was showing regular evolution towards the attractor which slowed
in proportion to the amount of unbound material.

We conclude from this that neither RI nor ROI are driving the convergence as repeated expansions still lead to the attractor. This is in agreement with \citet{HansenOct} where different perturbations
were presented, all of which lead to the attractor even in cases where the structures remained
perfectly spherical throughout.

\section{Testing the requirements}
Having demonstrated that ROI is not the driving force behind the attractor
we now consider possible origins discussed in the literature, 
namely energy exchange and phase mixing in a dynamic potential. We will show that 
those two effects are both necessary and sufficient conditions for the
emergence of the attractor. Since these effects are always present during
mergers this shows the potential importance of the attractor
for cosmological collisionless structures \citep{Ludlow2010}.

Energy exchange refers to energy being passed between particles by gravitational interaction. While it may seem that any
evolving system must exchange energy to evolve, it is possible to
design perturbations that change the system without exchanging energy
between particles. For example, one could design a kick which moved
the system in phase space by rotating velocity vectors, which would
perturb the system but would not cause the particles to exchange
energy; they would all still be on stable orbits since their kinetic and potential energies are identical
to their previous ones, hence the radial regions indicated by the
  apocenters of the particles barely change while the pericenters move
  because of kicks of the angular momentum.

\subsection{Energy exchange: the velocity anisotropy axis kick}
In order to investigate the importance of energy exchange between collisionless
particles, we will now construct a kick which conserves the energy
of each particle.

The new kick, which we will refer to as the `anisotropy kick', kicks
in velocity isotropy only. It aims to move the system in $\beta$ by rotating
each particle's velocity vector by a calculated amount. This does not
change the total energy in the bin - each particle independently and
exactly conserves its kinetic and potential energy - but does, by definition, change the angular momentum. The means by which the velocity rotations are performed is outlined in
appendix \ref{sec:construct}. The foundation of the method remains the same as from previous examples; alternating patterns of kick and flow, only now the kick is a function that rotates velocity vectors rather than randomly scaling individual components of velocity.

The resulting system will be slightly Radially Jeans
Unstable (RJU), i.e. not satisfying the static spherical Jeans equation, after the kick so the system will still need to re-establish
equilibrium. 
This kick should require the system to find a new equilibrium but does so without a prescribed way that the new equilibrium is
reached.

\begin{figure}
\includegraphics[width=84mm]{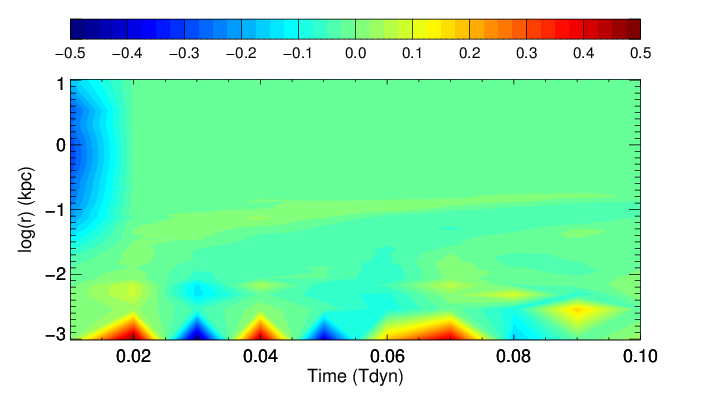}
\caption{\label{fig:betachange1} Contours showing the changes in
 velocity anisotropy of an initially radially anisotropic system as
 it recovers from a moderately isotropising kick. As the system
 relaxes over time after being kicked, it is measured every $1\%$ of
 a dynamical time and the velocity anisotropy of each bin is compared
 to its previous anisotropy. The change in anisotropy at a certain
 radius from moment to moment is represented by the colour of the
 contours, where blue colours indicate an ongoing change towards
 towards tangential anisotropy and red colours indicate likewise for
 radial anisotropy. Green represents no evolution of anisotropy between outputs.}
\end{figure}
\begin{figure}
\includegraphics[width=84mm]{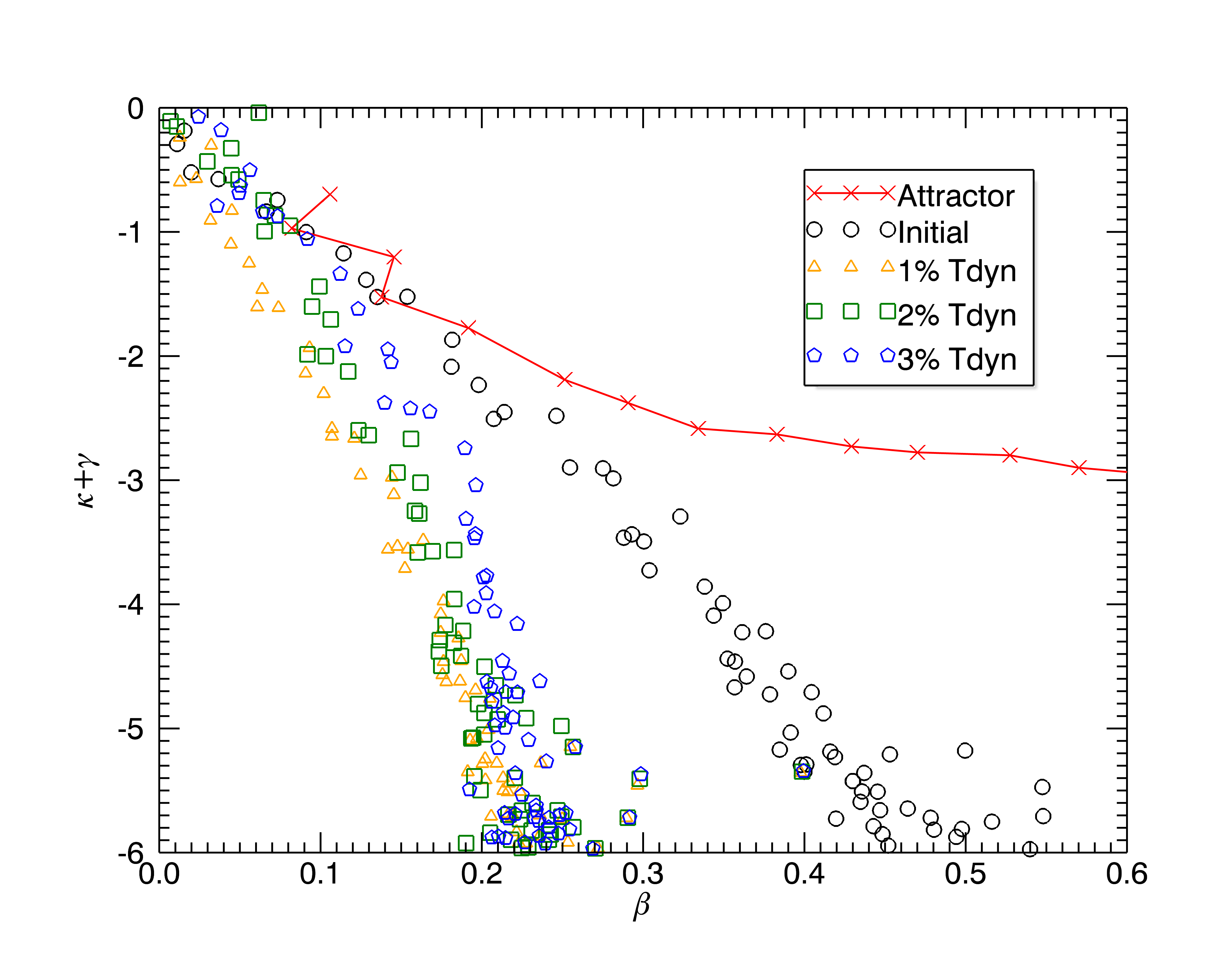}
\caption{\label{fig:attractortime} A more familiar presentation of some of the datasets represented in figure \ref{fig:betachange1} using the plot axes of the attractor space. Note the gentle
 drift of the data towards radial anisotropy during the first
 $3\%$ of a dynamical time after the kick, as summarised in figure
 \ref{fig:betachange1} by the yellow-green hue of the majority of the
 contours.}
\end{figure}

We first take an initial system with a radially anisotropic velocity
ellipsoid and force it
to become more isotropic. Figure \ref{fig:betachange1} examines the
change in the velocity anisotropy of the system as it equilibrates
after the kick. We define the `change in velocity anisotropy'
simply as $\frac{\Delta\beta}{\Delta t}$ where $\Delta t$ is time
between outputs of the state of the system \emph{i.e.}
$0.01T_{dyn}$. The kick is visible as the large, dark
blue section at the beginning of the time series. Afterwards, the
system relaxes over the course of about a dynamical time. The
relaxation is visible in the yellow-green tint across the rest of
plot, showing a general trend for the system to drift towards a more
radial velocity anisotropy again. See figure \ref{fig:attractortime}
for an alternative representation of some of the information displayed in figure
\ref{fig:betachange1}.

There are two particularly prominent features in figure
\ref{fig:betachange1} that require comment. The first is the kick itself,
clearly visible as a large, dark area along the left side, and
the other is the `sawtooth' pattern of spikes in velocity
anisotropy rate that run along the bottom of the graph. Both of these
features overshadow the actual point of interest, the general trend of
the plot, and accordingly all future plots will be cropped and re-scaled to present the clearest view of the
data. We will now take a moment to justify the removal of these features.

Firstly, removing the kick is regrettable but it is of such greater
magnitude than anything else in the plot that retaining it
over-saturates the important contours. The only useful information
that it contained was the colour (direction) of the kick
which will always be indicated.

Secondly, the `sawtooth' pattern that appears at very small radii is
caused by an unfortunate combination of two factors: the
logarithmic scale artificially overemphasising the relative importance
of the inner bins (in terms of how much of the contour area they
occupy), and the tendency for the very innermost bins to have an
extremely noisy velocity anisotropy as a result of the data analysis. Clipping those few bins cleans the data
considerably, removes only a small amount of particles, and does not destroy any useful information.

\begin{figure}
\includegraphics[width=84mm]{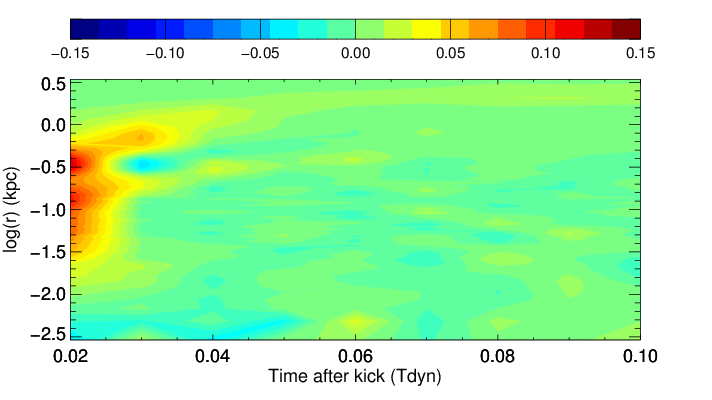}
\caption{\label{fig:betachange3} Contours showing the changes in
 velocity anisotropy of an initially radially anisotropic system as
 it recovers from a kick that set it to be completely
 isotropic. Notice that the system recovers by settling back towards
 radial anisotropy like figure \ref{fig:betachange1}, only much more
 strongly. Also note that, in line with the discussion in the body text and in contrast to figure \ref{fig:betachange1}, the oversaturated noise and kick features have been removed. Thus, the reddish spot along the left is not the kick, but the resettling of the system after the kick has occurred.}
\end{figure}

The amount of settling is a negative feedback effect that is a fraction of the size of the
perturbation. For example, if the system is initially strongly radial
and the kick is strong enough to make the system isotropic
then the settling will act to reverse the kick by drifting towards radial states, as seen in
figure \ref{fig:betachange3}. This drifting is stronger and more pronounced the larger the initial kick and is never enough to undo the kick.

\begin{figure}
\includegraphics[width=84mm]{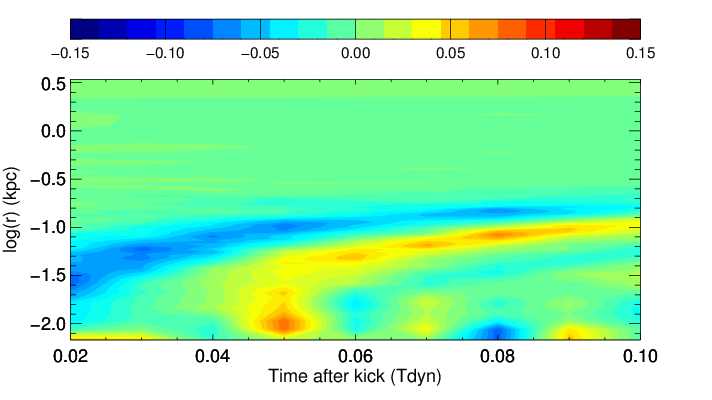}
\caption{\label{fig:betachange2} Contours showing the changes in
 velocity anisotropy of an initially tangentially anisotropic system
 as it recovers from an isotropic kick. The green-blue hue of most of
 the plot, demonstrating motion towards tangential anisotropy, shows
 that settling is not directed towards the attractor. The two large stripes across the plot are the equivalent of the dark spot in figure \ref{fig:betachange3} and are the settling of the system against the kick. Here the settling is towards a more tangential anisotropy and occurs as a ripple of anisotropy through the system from the inner regions to the outer. This effect is still minor compared to the kick and overall does still obey the established rule.}
\end{figure}

We will now demonstrate that this negative feedback
is not related to the attractor.
In figure \ref{fig:betachange2}
we take an initially tangentially anisotropic system and perturb it
towards isotropy. What we see is the reverse of figure
\ref{fig:betachange1} with the settling being more tangential and thus a light
blue.

This demonstrates that when a system is perturbed using
the anisotropic kick its subsequent relaxation will undo a small
fraction of that isotropy change. This means that 
while this perturbation does
destabilise the system and allow it to find new equilibria, it does
not lead towards the attractor.

In our final test of this, we took an initially slightly radially
anisotropic system and repeatedly perturbed it such that the system gradually moved
up to, and then past, the phase-space region of the attractor. The reasoning behind this scheme
is that if the system is still affected by the attractor then we would
expect it to behave differently when it is passing over it, perhaps changing the magnitude or direction of the settling.

We found that none of the kick-flow cycles in the series showed any evidence of being drawn to the attractor. The velocity
anisotropy evolution remained comparatively featureless throughout and showed no behaviour different from any other system perturbed by the anisotropic kick.

Overall, it appears that this method of perturbing the system does
cause the system to undergo some slight evolution in response to the
kick, but it seems restricted to a weak, negative feedback effect that
bears no relation or correlation to the attractor.

\subsection{Phase mixing in a fixed potential: the massless kick}
By phase mixing we mean 
how particles disperse through the phase space of the system, generally reducing the coarse grained
phase-space density by filling their orbital tori evenly. This is a kinematic process that causes a dispersion of particles along their orbits which even occurs in static potentials \citep{BinneyBook} and corresponds to the processes that occur during the `flow' periods of our perturbation schemes. In B12 we 
showed that repeated kicks, without subsequent periods of settling flow, do
not lead to the attractor. Here we will further emphasise the importance of the dynamics
of the flow by considering relaxation in a fixed potential.

This perturbation involved making the particles
massless. We took the same initial plummer spheres as before but then
froze the system's numerical, not analytical, potential and
transformed the particles into a population of massless tracer
particles. This means the background potential is no longer coupled to the particle distribution and,
because the simulation is collisionless, the particles have no way of
interacting with each other. After the kick has
occurred the particles will not be able to directly influence each others
positions in phase space.

If the attractor is driven only by the kick then removing the dynamical potential should have minimal effect on the system's convergence to
the attractor as simply the act of kicking would cause convergence.

\begin{figure}
\includegraphics[width=84mm]{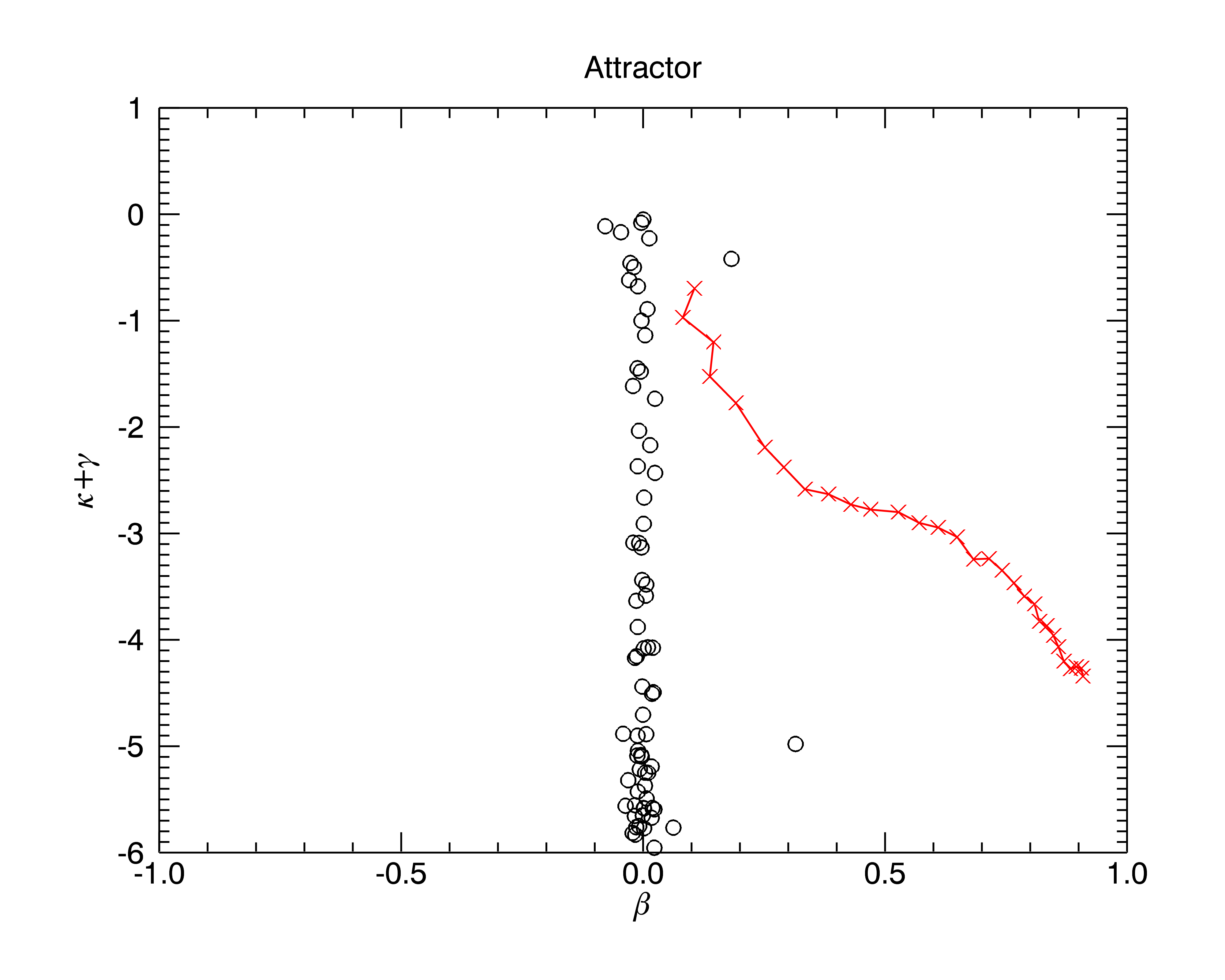}
\caption{\label{fig:phase} A plot showing an initially anisotropic
 system after an application of the `massless kick'. Comparison
 between the current system and its initial state is made difficult
 as the system has not evolved at all as a result of the kick. The black circles are are bins of the system after one kick
and were completely unaffected by the kick. The red line is the attractor.}
\end{figure}

We show the effects of applying the normal
scaling kick of B12 to our massless system in figure \ref{fig:phase}. The system has not evolved since
the kick and has certainly not moved towards the
attractor. This shows that the attractor does
not arise from the statistics of the kick mechanism alone
and does in fact require the subsequent mixing in a dynamical potential. 
Such mixing in a dynamical potential is fortunately always present during realistic 
cosmological structure formation.

\section{Summary}
We have addressed the fundamental physical mechanism responsible for the attractor.
We find that the radial orbit instability is not the underlying reason
for the robustness of the attractor.

Instead, we find that both energy exchange and phase mixing in a dynamical potential are necessary
conditions for the appearance of the attractor. Since earlier studies have
indicated that those two are sufficient conditions \citep{HansenMay, HansenOct},
 we believe we have established the physics underlying the attractor.
Since both effects are present during structure formation, in particular
through violent relaxation of mergers, this shows that the attractor
is relevant for the fully equilibrated part of cosmological structures.

\section{Acknowledgements}
The authors gratefully acknowledge the invaluable help and input of
Xufen Wu of the University of Bonn.
The Dark Cosmology Centre is funded by the Danish National Research Foundation.
This research was funded in part by the Science and Technology Facilities Council.

\bibliographystyle{mn2e}
\bibliography{attractor_paper2}

\vfill \eject

\appendix

\section{Construction of the isotropising kick}
\label{sec:construct}

\begin{figure}
\includegraphics[width=84mm]{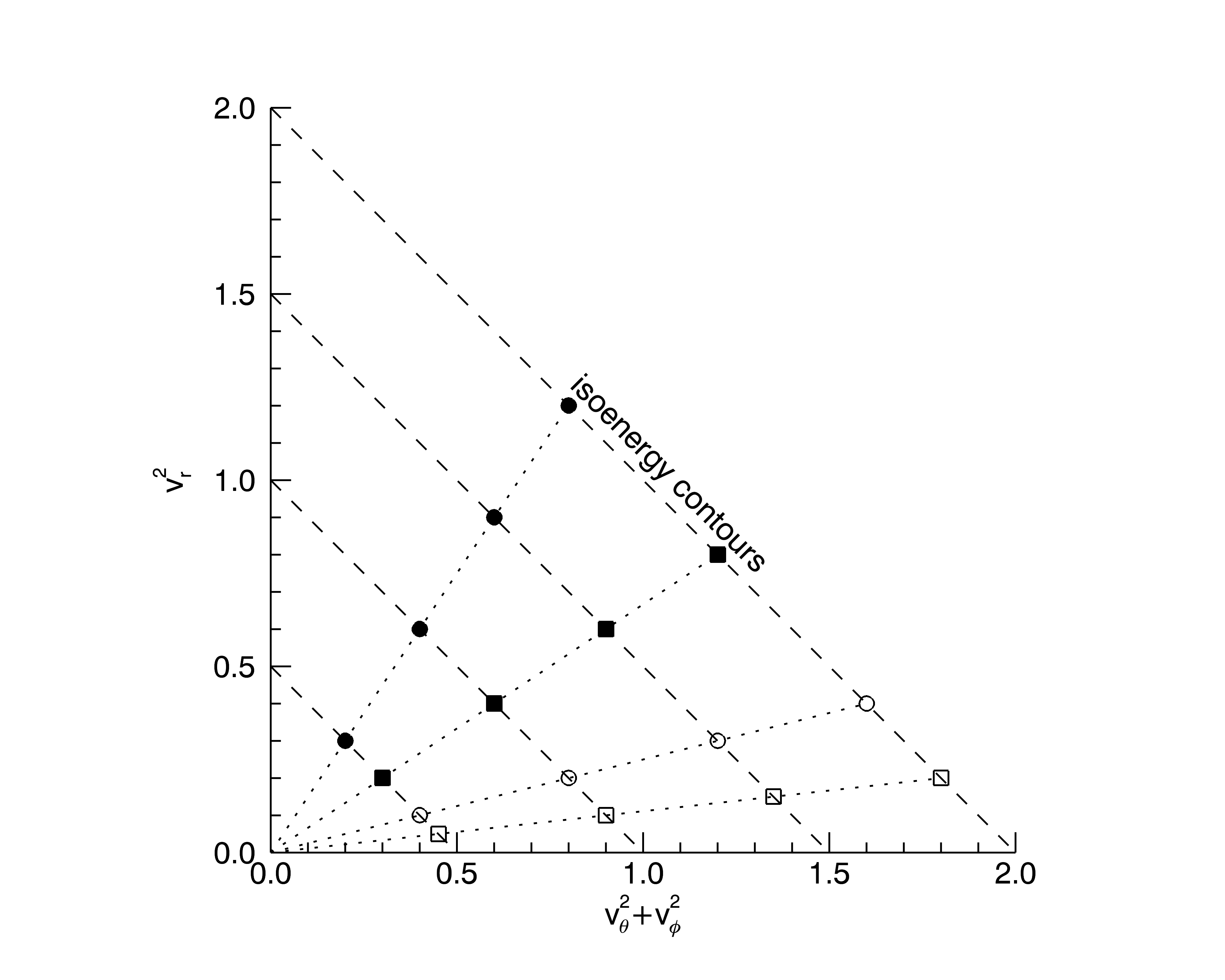}
\caption{\label{fig:schematic} A visual representation of how the perturbation changes anisotropy. Particles are moved along isoenergy contours from the open symbols to the closed symbols. This plot shows a kick of $\alpha=6$ applied to two groups of particles, one with most of their energy in the tangential velocity components (squares) and the other with a more even distribution (circles). The distance moved along the isoenergy contours depends on the particle's initial position along them.}
\end{figure}

The nomenclature used in this section is as follows: A bin in our
system has a population of $n$ particles that give the bin an
anisotropy of $\beta$ based on their kinetic energy, $T$, in the
radial, $T_r$, and tangential, $T_t$, directions; $T_t$ being made up
of $T_\theta$ and $T_\phi$. We are talking about a perturbation, so we
speak in terms of an \emph{initial} state, $\beta_1$, and a
\emph{final} state, $\beta_2$. We find it helpful to specifically define $T_t$ as $\frac{1}{2}(T_\theta + T_\phi)$ as this simplifies matters.

As such, we begin from:

\begin{equation}
\beta_1=1-\frac{\sum\limits_{i=1}^n{T_{t1}}_i}{\sum\limits_{i=1}^n{T_{r1}}_i}
\end{equation}

Our perturbation acts to move the anisotropy from $\beta_1$ to
$\beta_2$, the change being expressed as: $x\beta_1=\beta_2$, so we
can say:

\begin{equation}
\label{eqn:a1}
x\beta_1=1-\frac{\sum\limits_{i=1}^n{T_{t2}}_i}{\sum\limits_{i=1}^n{T_{r2}}_i}=1-\frac{a(x)\sum\limits_{i=1}^n{T_{t1}}_i}{b(x)\sum\limits_{i=1}^n{T_{r1}}_i}
\end{equation}

where $a$ and $b$ are just another, more helpful way of assessing the
impact of $x$ on the particle energies. Speaking of the particle
energies, we require global energy conservation, so we specify that:

\begin{equation}
 2a(x)\sum\limits_{i=1}^n{T_{t1}}_i+b(x)\sum\limits_{i=1}^n{T_{r1}}_i=2\sum\limits_{i=1}^n{T_{t1}}_i+\sum\limits_{i=1}^n{T_{r1}}_i=\mathcal{E}
\end{equation}

where $\mathcal{E}$ is the system's overall kinetic energy. We can
thus create definitions of $a$ and $b$,

\begin{equation}
a(x)=\frac{\mathcal{E}-b(x)\sum\limits_{i=1}^n{T_{r1}}_i}{2\sum\limits_{i=1}^n{T_{t1}}_i}\text{; }b(x)=\frac{\mathcal{E}-2a(x)\sum\limits_{i=1}^n{T_{t1}}_i}{\sum\limits_{i=1}^n{T_{r1}}_i}
\end{equation}

and feed them into each other to get solutions that are still linked
by energy conservation but can be expressed separably;

\begin{equation}
a(x)=\frac{\mathcal{E}}{\sum\limits_{i=1}^n{T_{t1}}_i\left[2+\frac{1}{1-x\beta_1}\right]}\text{; }b(x)=\frac{\mathcal{E}}{\sum\limits_{i=1}^n{T_{r1}}_i\left[3-2x\beta_1\right]}
\end{equation}

This tells us how the bin as a whole must act, but does not tell us
how to achieve this by manipulating individual particles. To move on
to that stage we must make $a$ and $b$ more applicable to each
particle.

When we scale the tangential energy, $\sum\limits_{i=1}^n{T_{t1}}_i$,
by $a$, what we are actually doing is multiplying each particle's
energy by some number, quite possibly a \emph{different number for
 each of them}, and we need a way to determine what that number
should be. To that end, we create two numbers, $d$ and $e$, and let them take
different values for each particle, $i$.

For convenience we do not write out the dependence of $d$ and $e$ on $a$
\emph{vis.} $d(a)_i$. This is primarily to reduce clutter and because
the final result of the process will not need to refer to $a$, $b$, or
any of the other scale factors introduced in this process.

\begin{equation}
\label{eqn:a2}
\sum\limits_{i=1}^n{T_{t2}}_i=a(x)\sum\limits_{i=1}^n{T_{t1}}_i=\sum\limits_{i=1}^nd_i{T_{t1}}_i=\frac{\mathcal{E}}{1+\frac{1}{1-x\beta_1}}
\end{equation}

\begin{equation}
\label{eqn:a3}
\sum\limits_{i=1}^n{T_{r2}}_i=b(x)\sum\limits_{i=1}^n{T_{r1}}_i=\sum\limits_{i=1}^ne_i{T_{r1}}_i=\frac{\mathcal{E}}{2-x\beta_1}
\end{equation}

As well as global energy conservation, we now specify that we would
like energy conservation at the particle level as well:

\begin{equation}
2d_i{T_{t1}}_i+e_i{T_{r1}}_i=2{T_{t2}}_i+{T_{r2}}_i=E_i
\end{equation}

At this point is is most convenient to start constructing the problem
in terms of a single variable that we must solve for, $\alpha$:

\begin{equation}
\frac{{T_{t2}}_i}{{T_{r2}}_i}=\frac{d_i{T_{t1}}_i}{e_i{T_{r1}}_i}=\alpha_i\frac{{T_{t1}}_i}{{T_{r1}}_i}
\end{equation}

This allows energy conservation to be rephrased in ways such as:

\begin{equation}
\left[2\alpha_i\frac{{T_{t1}}_i}{{T_{r1}}_i}+1\right]e_i{T_{r1}}_i=E_i
\end{equation}

By manipulating energy conservation in this way, we arrive at the definitions:

\begin{equation}
d_i{T_{t1}}_i=\frac{E_i}{2\alpha_i\frac{{T_{t1}}_i}{{T_{r1}}_i}+1}\alpha_i\frac{{T_{t1}}_i}{{T_{r1}}_i}
\end{equation}

and:

\begin{equation}
e_i{T_{r1}}_i=\frac{E_i}{2\alpha_i\frac{{T_{t1}}_i}{{T_{r1}}_i}+1}
\end{equation}

By taking these results back to equations \ref{eqn:a2} and
\ref{eqn:a3} and then combining them with our starting point of
equation \ref{eqn:a1}, we arrive at our final result:

\begin{equation}
\label{eqn:refer}
\frac{\sum\limits_{i=1}^n\frac{E_i}{2\alpha_i\frac{{T_{t1}}_i}{{T_{r1}}_i}+1}\alpha_i\frac{{T_{t1}}_i}{{T_{r1}}_i}}{\sum\limits_{i=1}^n\frac{E_i}{2\alpha_i\frac{{T_{t1}}_i}{{T_{r1}}_i}+1}}=1-x\beta_1
\end{equation}

There are a lot of solutions sets for $\alpha$ that will yield the
result we want and we have no way of choosing between them without
stating another condition. The condition we set is that $\alpha$ has
one fixed value for each mass bin, and then we solve the equation
iteratively.

\bsp

\label{lastpage}

\end{document}